\documentclass[structabstract]{aa}
\usepackage{txfonts}
\usepackage{natbib}
\bibpunct{(}{)}{;}{a}{}{,} % to follow the A&A style
\usepackage{graphicx}
\inputencoding{utf8}
\begin{document}

\title{Temperatures and metallicities of  M giants in the galactic Bulge  from low-resolution K-band spectra \thanks{Based on observations collected at the European Southern Observatory, Chile, program number 089.B-0312B}
}

%\author{M. Schultheis \& B.Q. Chen et al.}

\author{M. Schultheis \inst{1}
\and   N. Ryde \inst{2}
\and G. Nandakumar \inst{1}
}

   \institute{ Laboratoire Lagrange, Universit\'e C\^ote d'Azur, Observatoire de la C\^ote d'Azur, CNRS, Blvd de l'Observatoire, F-06304 Nice, France
 e-mail: mathias.schultheis@oca.eu
 \and
 Department of Astronomy and Theoretical Physics, Lund Observatory, Lund University, Box 43, 221 00, Lund, Sweden e-mail:ryde@astro.lu.se
 }

\abstract {With the existing and upcoming large multi-fibre low-resolution spectrographs, the question arises how precise stellar parameters such as $\rm T_{eff}$ and [Fe/H] can be obtained from low-resolution K-band spectra with respect  to traditional photometric temperature measurements. Until now, most of the effective temperatures in galactic Bulge studies come directly from  photometric techniques. Uncertainties in interstellar reddening and in the assumed extinction law could lead to large systematic errors ($\rm > 200\,K$). }
 {We aim to obtain and calibrate the relation  between $\rm T_{eff}$ and the $\rm ^{12}CO$ first overtone bands for M giants in the galactic Bulge covering a wide range in metallicity.  }
{ We use low-resolution spectra for 20 M giants with well-studied  parameters from photometric measurements   covering the temperature range $\rm 3200 < T_{eff} < 4500\,K$ and a metallicity range from 0.5\,dex down
to -1.2\,dex and study the behaviour of $\rm T_{eff}$  and [Fe/H] on the spectral indices.}
{ We find a tight relation  between $\rm T_{eff}$ and the $\rm ^{12}CO(2-0)$   band with a dispersion of 95\,K as well as between $\rm T_{eff}$ and the $\rm ^{12}CO(3-1)$ with a dispersion of 120\,K.   We do not find any dependence of these relations on
the metallicity of the star, making them relation attractive for galactic Bulge studies. This relation is also not sensitive to the spectral resolution allowing to apply this relation in a more  general way.
 We also found a  correlation between  the  combination of the NaI, CaI and the $\rm ^{12}CO$   band  with the metallicity of the star. However this relation is only valid for sub-solar metallicities. }
{ We show that low-resolution spectra provide a powerful tool to obtain effective temperatures of M giants. We  show that this relation does not depend on the metallicity of the star within the investigated range and is also applicable to different spectral resolution making this relation in general useable to derive effective temperatures in high extincted regions where photometric temperatures are not reliable. }


\maketitle

\titlerunning{Low resolution K-band spectra}
\authorrunning{Schultheis \& Ryde}

\section{Introduction}

%With the upcoming large surveys (APOGEE-2,GALAH,Gaia-ESO,LAMOST,WEAVE, etc..) detailed chemical abundances  of a large amount of stars will be provided to the community. Efficient and accurate effective temperature  determination is  essential to get reliable abundances. 
 While photometric temperatures can be estimated quite precisely in low-extinction windows (see e.g. \citealt{hernandez2009}),  regions where interstellar reddening is high (e.g. galactic Bulge, galactic plane), photometric temperatures, by definition,  suffer from large and unknown systematic uncertainties. Going even closer to the Galactic Centre region ($\rm R_{GC} < 200\,pc$) it is virtually impossible to get reliable photometric temperatures due to the extreme high interstellar reddening (\citealt{schultheis2009}, \citealt{gonzalez2012}, \citealt{schultheis2014}).  To overcome this problems, \citet{ryde2015}, for example, instead derived $\rm T_{eff}$ from spectral indices in low-resolution K band spectra in their study of Galactic center stars. They were then able to obtain accurate detailed chemical abundances of nine M giant stars close to the Galactic Center using the CRIRES high-resolution spectrograph.
This technique  was extended to latitudes at $\rm b=-1^{o}$ and $\rm b=-2^{o}$  (\citealt{ryde2015a}).

\citet{ramirez1997} and \citet{ramirez2000}  studied the behaviour of the $^{12}$CO band head situated at 2.3\,$\mu$m with low-resolution K-band spectra ($\rm R \sim 2000-4000$) and found for M giants a remarkably  tight relation between the equivalent width (EW) and the effective temperature. \citet{blum2003}  and \citet{ivanov2004} confirmed this strong temperature dependence of the $^{12}$CO band head using different spectral resolution. 
%\citet{pfuhl11} used integral field spectroscopy of 450 cool giants within  1\,pc from the GC to derive the star formation history of the Milky Way. They derived effective temperatures from the CO bandheads.

\citet{ramirez2000}, \citet{frogel2001},  \citet{schultheis2003} and \citet{ivanov2004} found  that the combined index of the Na{\sc i} doublet at 2.21\,$\mu$m  and the Ca{\sc i} triplet at 2.261\,$\mu$m is very sensitive to the surface gravity of the star  and can be used to separate, e.g., M giants from supergiants or dwarf stars.  \citet{schultheis2003} has shown the power of low-resolution spectra in the region of high interstellar extinction. They could separate different stellar populations such as red giant branch stars, AGB stars, supergiants or young stellar objects.
\citet{frogel2001} and \citet{ramirez2000}   obtained relations between EW(Na{\sc i}), EW(Ca{\sc i})  and [Fe/H] which were calibrated on globular clusters. Based on this calibration,  \citet{ramirez2000} studied the metallicity distributions in the Galactic center region and  found no evidence of a metallicity gradient for the Inner Bulge.
\citet{pfuhl11} determined the average star formation rate from 450 cool giant stars located in the nuclear star cluster. They obtained low-resolution spectra ($\rm R \sim 2000-3000$) of 33 giants in the solar neighbourhood ($\rm -0.3 < [Fe/H] < 0.2$) with spectral types from G0--M7 and obtained a $\rm CO-T_{eff}$ relation with a residual scatter of 119\,K.  
However, the $\rm T_{eff}$ vs. $\rm ^{12}CO$  calibration has been only investigated for a bright local solar-neighbourhood sample with a narrow metallicity range.
 We extend this study for M giants located in the galactic Bulge  using a wide  metallicity range to test this calibration.

Until now most of the abundance studies (e.g. Rich \& Origlia 2012, Gonzalez et al. 2011, Zoccali et al. 2008) in galactic Bulge fields use
photometric temperatures based on a colour-temperature relation and assuming interstellar reddening values {\bf{as a first temperature estimate.}} However, variable extinction, uncertainties in the extinction law, etc.  can lead to  severe uncertainties  in the derived temperatures ($\rm > 200\,K$) which can lead to unknown and different systematic offsets, in the abundance determination. Here we apply and investigate these relations for stellar abundances studies in the Bulge, extending their use for a wide range of metallicities

%The galactic Bulge with its wide dispersion of metallicity is  the ideal laboratory to study  questions such as:
% How do derived temperatures, gravities and metallicities  compare between low-resolution and high-resolution IR spectra?   How do the temperatures compare when using classical photometric colours such as J--K? What is the influence of the $\alpha$-elements on the
%temperature-colour relations?

In this paper, we show how low-resolution spectra can provide a powerful tool to get high accurate effective temperatures
in highly extincted regions (such as the Galactic Center). We are able to show that this method chosen by, e.g.  \citet{ryde2015} and \citet{ryde2015a}, indeed is a good choice. 
The paper is structured as following: In Sect. \ref{observations} we describe the data and the data reduction
process, in Sect. 3 we discuss how the known effective temperature,  and metallicity of our calibration stars relates with the CO bands and the spectral indices such as  NaI and CaI, in Sect. 4 we apply our method to M giants in the inner Galactic bulge  and we finish in Sect. 5 with the conclusions.

\section{Observations} \label{observations}

We obtained near-IR spectra on 27 June -- 30 June 2010 with ISAAC at ESO, Paranal, Chile. We used the red grism of the ISAAC spectrograph, covering 2.0--2.53 $\mu$m, to observe 21 M giants in the galactic Bulge. We took the spectra 
 under photometric conditions through a 1\,$\arcsec$ slit providing a resolving power of  R $\sim$ 2000.
We obtained a Ks-band acquisition image before each spectrum  to identify the source and place it on the slit.
We used UKIDSS finding charts for source identification and to choose ``empty''  sky positions for sky subtraction
for each source along the 90\,$\arcsec$ slit. We used an ``ABBA'' observing sequence for optimal sky subtraction.

We observed B dwarfs (typically 6-8 stars per night), close to the airmass range of our targets, as telluric standard stars to correct for the instrumental and atmospheric transmission. We used {\tt IRAF} to reduce the ISAAC spectra.
We removed cosmic ray events, subtracted the bias level, and then divided all frames
by a normalized flat-field.  We used the traces of stars at two different (AB) positions
along the slit to subtract the sky.  After extracting and co-adding the spectra, we calibrated wavelengths using the Xe-lamp. The r.m.s of the wavelength calibration is better than 0.5 \AA.

We rebinned the spectra to a linear scale, with a dispersion of $\sim$ 7\,$\AA\
$ pixel and a wavelength range from 2.0 $\mu$m to 2.51 $\mu$m. 
We then divided each spectrum by the telluric standard observed closest in time and in airmass (airmass difference $<$ 0.05). We then normalized the resulting spectra by the mean flux
between 2.27 and 2.29 $\mu$m.   Table\ref{log} shows the list of known
Bulge stars with precise stellar parameters. Our stars lie in the temperature range $\rm 3200 < T_{eff} < 4500\,K$ and are selected from \citet{rich2005}, \citet{rich2012}, \citet{gonzalez2011a} and \citet{monaco2011}. They span a wide range
in metallicities ensuring that we can study an eventual  metallicity dependance of the CO vs. $\rm T_{eff}$  relation

\begin{figure}[!htbp]
\includegraphics[width=0.5\textwidth]{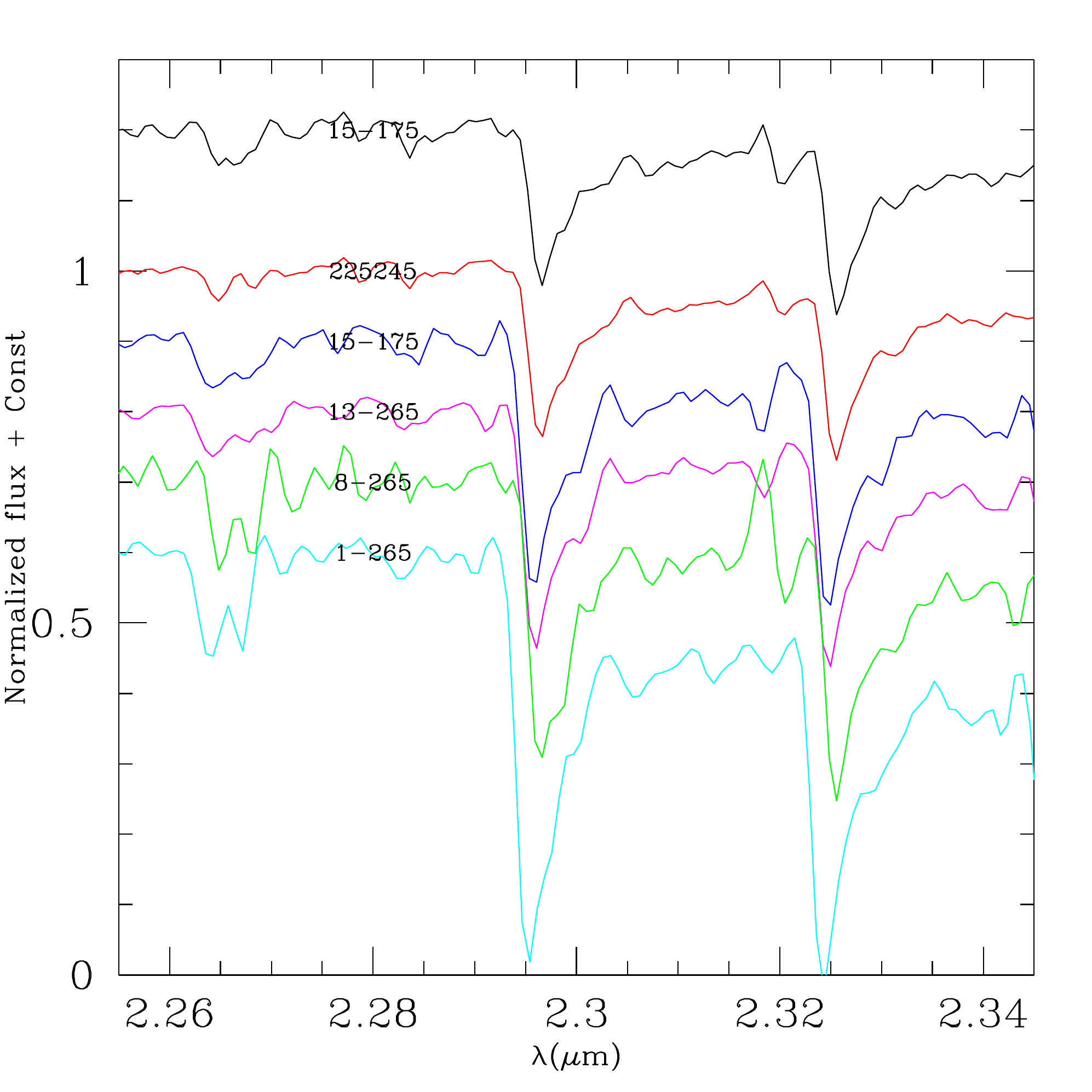}
\caption{Temperature sequence of the CO bandhead starting from 4500\,K (black) going down to 3200\,K (cyan) in 200\,K steps. Indicated are the star names (see Table. 1). }
\label{specobs}
\end{figure}

Figure~\ref{specobs} shows the temperature sensitivity of the CO bandhead where we see a steady increase of the absorption band with effective temperature.

\subsection{Stellar samples: calibrators and target M giants}
Our sample of stars cover  M giants sample  in  Baade's window from \citet{rich2005} and \citet{gonzalez2011}. All are from low extinction fields where
we can trust and account for the reddening in an accurate way.
In addition, we include stars
of \citet{rich2012} which are located at  $\rm l= 0^{o}$ and $\rm b=-1^{o}$. This field which is close to the Galactic Center shows well studied  interstellar extinction  (see e.g. \citealt{schultheis1999}, \citet{gonzalez2012},etc..).
The other stars come from the thick disc study of \citet{monaco2011} where the temperatures were obtained by using the dereddened $\rm (J-K)_{0}$ colour but with small and well quantified reddening. 

We will use our derived calibration based on the stars in Table 1,  on a sample of stars in the highly extincted inner bulge region. Thus, we have also observed,  with the
same instrument setup 28 Bulge M giants within 2 degrees from the Galactic Center, along the Southern minor axis to obtain effective temperatures.
 These 28 stars were also observed with CRIRES in order to get detailed chemical abundances (see \citealt{ryde2015}, \citealt{ryde2015a}). 9 stars were observed in the Galactic Center (\citealt{ryde2015}),  9 stars at $(l,b) = (0,-1^{\circ})$, and 10 stars at  $(l,b) = (0,-2^{\circ})$. These stars are M giants with  $\rm T_{eff}$=3300-4200K  and  $\rm 0.7  < log g < 2.25$, for which metallicities, [Fe/H], and the abundances of the $\alpha$ elements Mg, Si, and Ca were determined. Table~\ref{nils} shows the M giant sample of \citet{ryde2015a} together with  the derived effective temperatures and metallicities in this work (see Sect. 4) and the  metallicities derived in \citet{ryde2015a}.

\begin{table*}
\caption{List of observed targets together with ra, dec, K magnitude, spectral type, $\rm T_{eff}$, log\,g and [Fe/H]. }
\begin{tabular}{cccccccc}
name&ra&dec&K&Sptype&Teff& log\,g& [Fe/H]\\
\hline
BMB28$^{1}$&18:02:59.51&-30:02:54.3&7.39&M7&3400&0.5&-0.22\\
BMB55$^{1}$&18:03:08.11&-29:57:48.0&7.78&M8&3200&0.5&-0.17\\
BMB93$^{1}$&18:03:22.34&-30:02:56.4&7.28&M6&3600&0.5&-0.15\\ 
BMB124$^{1}$&18:03:29.72&-29:55:57.6&8.84&M6&3600&0.5&-0.15\\
BMB152$^{1}$&18:03:36.92&-30:01:47.4&7.31&M9&3200&0.5&-0.24\\
BMB165$^{1}$&18:03:43.78&30:05:17.2&8.10&M7&3400&0.5&-0.29\\
BMB289$^{1}$&18:04:22.66&-29:54:51.8&6.20&M9&3200&0.5&-0.15\\
1-265$^{2}$&17:58:37.12&-29:03:48.4&8.34&M9&3200&--&-0.08\\ 
8-265$^{2}$&17:58:43.83&-29:07:42.5&8.23& M7&3400&--&-0.19\\
13-265$^{2}$&17:58:37.41&-28:59:33.2& 8.32&M6&3600&--&-0.36\\
142173$^{4}$&00:32:12.56&-38:34:02.3&9.29&--& 4330&1.50&-0.69\\
15-175$^{2}$&17:52:53.33&-29:58:49.8&8.90&M5&3800&--&-0.37\\
171877$^{4}$&00:39:20.23&-31:31:35.5&8.12&--&3930&1.10&-0.77\\ 
225245$^{4}$&00:54:46.38&-27:35:30.4& 8.79&--&3920&0.65&-0.99\\ 
313132$^{4}$&01:20:20.66&-34:09:54.1&7.04&--&4530&2.0&0.01\\
343555$^{4}$&01:29:42.01&-30:15:46.4&8.46&--&4530&2.25&-0.62\\ 
42$^{3}$&18:10:17.65&-31:45:38.9&8.81&--&3750&1.20&-0.96\\
6$^{3}$&18:09:59.53&-31:38:14.1&12.08&--&3900&1.47&0.20\\ 
86$^{3}$&18:35:15.24&-34:46:41.4&12.41&--&3850&1.66&0.42\\
BD-012971$^{1}$&14:38:48.04&-02:17:11.5&4.3&M5&3600&0.5&-0.78\\ 
%HD787$^{4}$&00:12:09.99&-17:56:17.8&1.855&--&3870 &--&--\\
\hline
\footnote{test} Rich\& Origlia (2005)\\
\footnote{test1} Rich et al. (2012)\\
\footnote{test2} Gonzalez et al. (2011)\\
\footnote{test3} Monaco et al. (2011)\\
\end{tabular}
%\caption{List of observed targets together with ra, dec, K magnitude, spectral type and effective temperature. }
\label{log}
\end{table*}

\begin{table}[!htbp]
\caption{Effective temperatures from this work, metallicities from the high resolution work of  \citet{ryde2015a},  and metallicities from this work (third column) of  the galactic Bulge M giant sample.}
\centering
\begin{tabular}{lccc}
\hline 
\hline
Star & $\rm T_{eff}$ & $\rm [Fe/H]_{CRIRES} $ &$\rm [Fe/H]_{low}$  \\
\hline
GC1  &3667&  0.15 	 & -0.15 \\   
GC20 &3684& 0.14 	 & -0.05\\   	
GC22 &3614& 0.04 	 & -0.02 \\ 
GC25 &3648 &-0.20        & 0.07 \\	  
GC27 &3405&  0.23 	 & 0.19 \\
GC28 &3736&-0.04        & -0.30\\ 
GC29 & 3420 & 0.12 &    -0.31 \\
GC37 &3747 & -0.08  & 0.19 	\\
GC44 & 3461& 0.18 	 & 0.24\\
\hline
bm1-06 &  3764&  0.29       &   -0.1   \\ 
bm1-07 &   3863& 0.08   &   -0.23    \\
bm1-08 &  3618 & 0.17    &  -0.25   \\
bm1-10 &3762 & -0.23    &   -0.45   \\
bm1-11 & 3734& 0.12     & -0.14  \\
bm1-13 & 3717& -0.94   &  -0.53   \\
bm1-17 &3770& -0.83     &  - 0.844   \\
bm1-18 &3781&   0.22    &  -0.28  \\
bm1-19 & 3905&  0.18     &   -0.17   \\
\hline
bm2-01  & 3948&  0.14  	 &  -0.62  \\
bm2-02  & 3990&-0.48  	  &  -0.60   \\
bm2-03  & 3666& 0.26  	  &  -0.34      \\   
bm2-05 & 3451& 0.01 	  &  -0.04     \\
bm2-06 & 4172& -1.17 	  &  -1.23     \\ 
bm2-11 &3961&-0.91 	  &  -0.85    \\
bm2-12 &3943& -0.11  	 &  -0.28    \\ 
bm2-13 &3746& -0.16  	  & -0.42  \\ 
bm2-15 & 3984& 0.22  	  & -0.13     \\      
bm2-16 &  3838&0.10  	 &  -0.16    \\
\hline

\end{tabular}
\label{nils}
\end{table}
\section{The method: empirical temperatures and metallicities} \label{Results}

We discuss in this Sect. our method  based on our calibration sample (Table.~1).  

\subsection{Effective temperature}

When measuring the equilavent width of the   $^{12}$CO band,  the choice of the pseudo-continuum bands is important (see e.g. \citet{ivanov2004}).  We have tested different continuum bands such as \citet{ramirez2000}, \citet{frogel2001}, \citet{ivanov2004}, \citet{pfuhl11} and looked for the smallest dispersion in $\rm T_{eff}$ in our sample. 
We have found that the \citet{blum2003} bandpasses and continuum points show in general the smallest r.m.s dispersion.  The absorption indices defined by Blum et al. (2003) are measured relative to spectral regions adjacent to the absorption itself and are thus independent of reddening. The $^{12}$CO(2-0)  index is defined as the percentage of the flux in the $^{12}$CO(2-0) feature relative to a continuum band centered at 2.284\,$\mu$m.  The $^{12}$CO(2-0) band and continuum band is 0.015\,$\mu$m wide and the $^{12}$CO(2-0) band is centered at 2.302\,$\mu$m (see \citealt{blum2003}).   In addition, we also  measure the  $^{12}$CO(3-1) bandhead.  Table 1 gives central wavelengths and bandpasses for the CO lines and their pseudo-continuum. 

\begin{table}
\caption{Band passes and continuum points of the $^{12}$CO bands.}
\begin{tabular}{lcc}
Feature&$\lambda_{C}$ (\AA)& $\Delta \lambda$ (\AA)\\
\hline
$^{12}$CO(2-0)& 23020 &150 \\
$^{12}$CO(2-0) continuum&22840  & 150\\
$^{12}$CO(3-1)& 23245& 54 \\
$^{12}$CO(3-1) continuum& 22840 & 150\\
\hline
\end{tabular}
\end{table}

\begin{figure}[!htbp]
\includegraphics[width=0.5\textwidth]{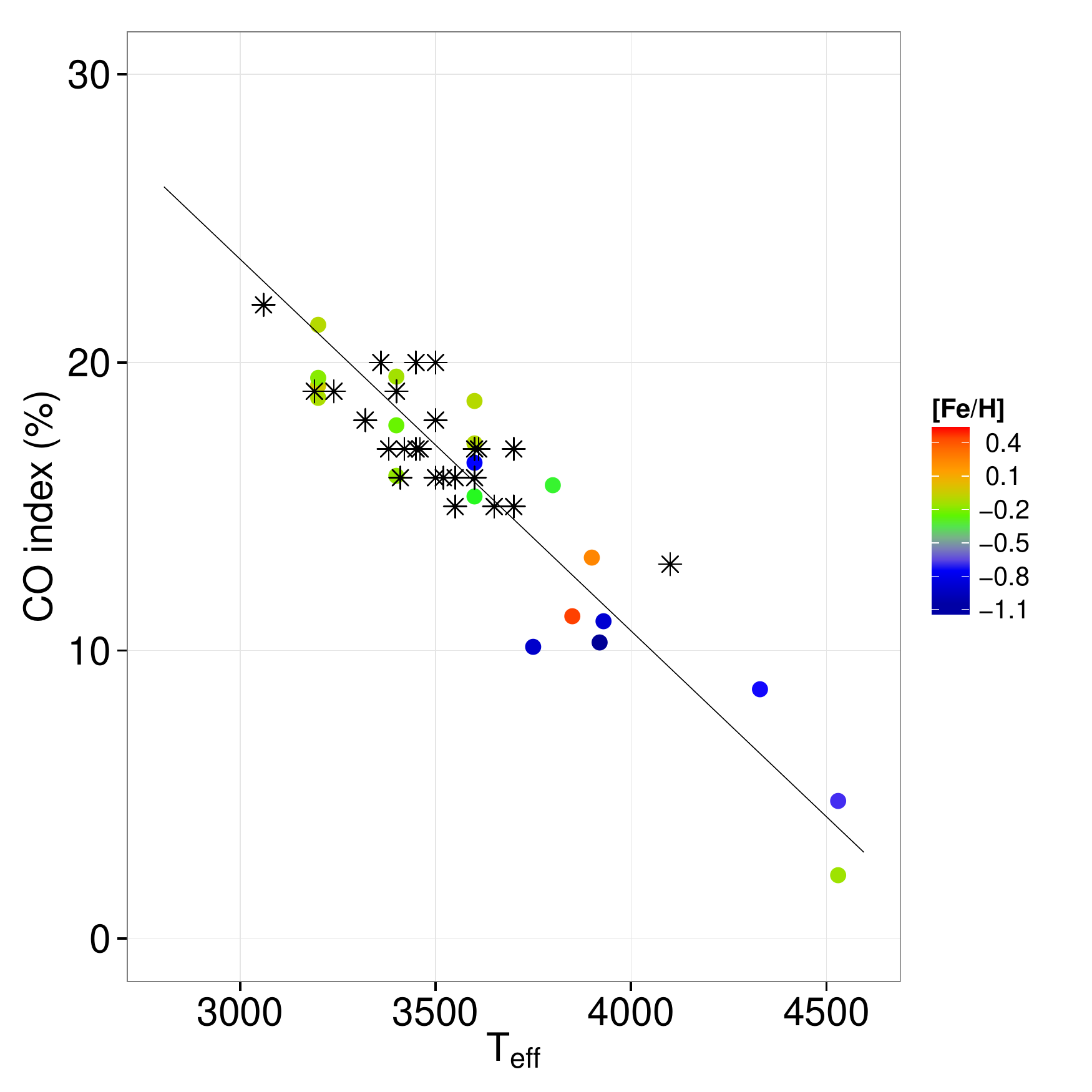}
\caption{Effective temperature vs.  $^{12}$CO(2-0)  band as a function of metallicity.  Black asterisks are the stars from Blum et al. (2003)  The straight line shows the fitted relation by Blum et al. (2003).}
\label{COvsTeffblum}
\end{figure}

Figure~\ref{COvsTeffblum} shows the $^{12}$CO(2-0) index of  the \citet{blum2003}  sample as well as our  sample. The black lines give the fitted relation of \citet{blum2003} with  {\bf{$\rm T_{eff} =4828.0 - 77.5 \times CO(2-0)$}}  which is very similar to our linear least-square fit. 
The r.m.s of the  fit is  95\,K which is comparable with the r.m.s scatter of \citet{pfuhl11}. 
% For stars hotter than 4000\,K the dispersion increases as the strength of the CO band decreases.
%\subsection{Effective temperature}

The Blum et al. (2003) sample covers bright solar neighbourhood giant stars (see their Table.3) with typical solar metallicities. We extended this work for
galactic Bulge stars in low extinction fields covering a wide metallicity range with $\rm -1.2 < [Fe/H] < 0.5$ which enables to test any possible metallicity dependence of this relation. As shown in Fig. ~\ref{COvsTeffblum}, we do not find any metallicity dependence on this relation within the metallicity range of our calibration stars, relevant for the bulge, making this method extremly interesting for galactic Bulge studies.  The spectral resolution of the comparison stars in
\citet{blum2003} is $\sim$ 750. As one sees in Fig.~\ref{COvsTeffblum} the effect of using a different instrument setup with a different spectral resolution (i.e. $\rm R \sim 2000$ vs. $\rm R\sim 750$) does not affect the 
 $^{12}$CO(2-0) vs. $\rm T_{eff}$ relation. This also indicates that the CO index is insensitive to spectral resolution and could therefore be used more generally.

\begin{figure}[!htbp]
\includegraphics[width=0.5\textwidth]{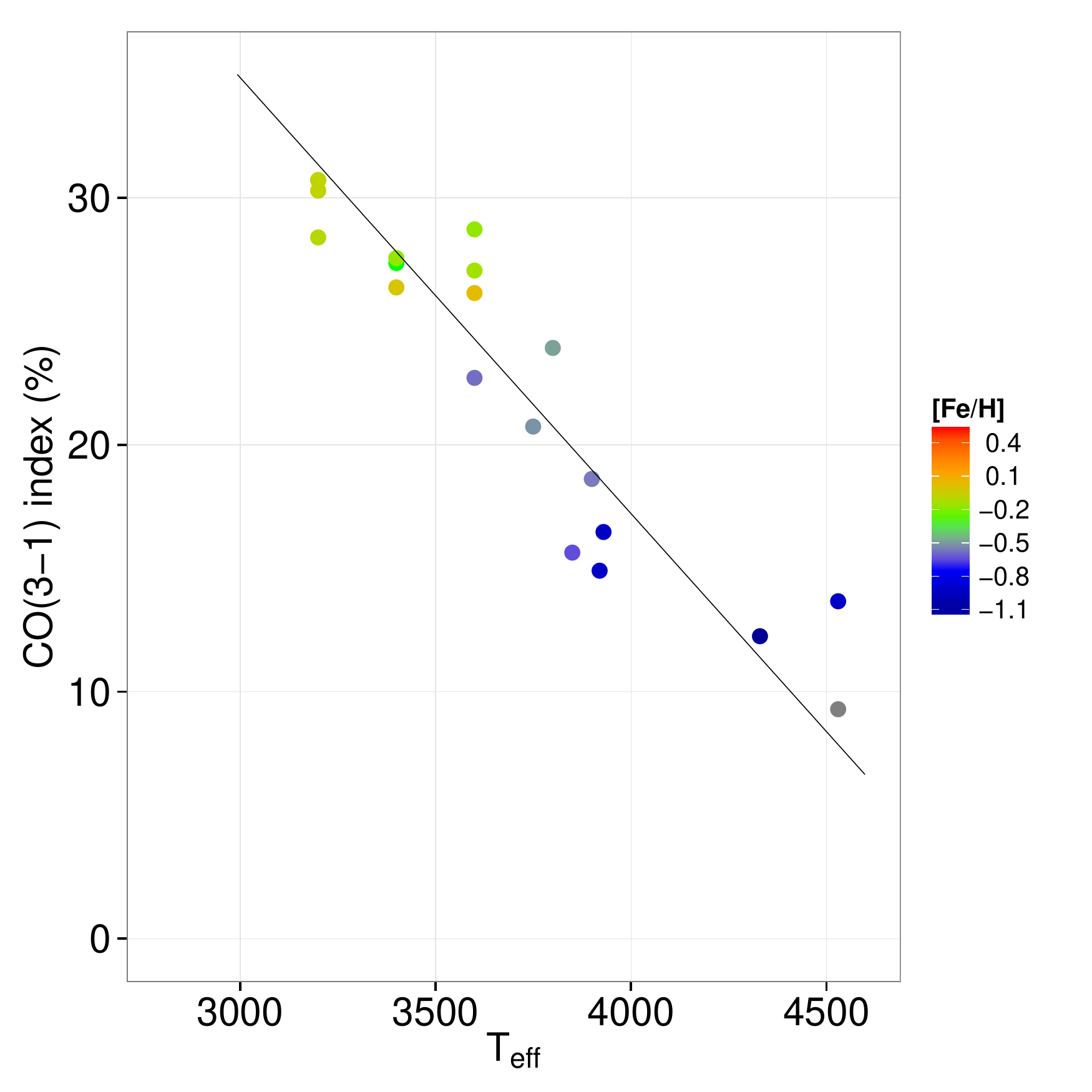}
\caption{Effective temperature vs.  $^{12}$CO(3-1)  band as a function of metallicity.   The straight line shows our best fit.}
\label{COvsTeffblum2}
\end{figure}

Figure~\ref{COvsTeffblum2} shows a similar plot but using the $^{12}$CO(3-1) band centered at $\rm \lambda_{c} = 2.3245\,\mu m$. We use the same continuum points as for the $^{12}$CO(2-0) band (see Table.~1). We see here again a very tight relation between $\rm T_{eff}$ and   the  \mbox{$^{12}$CO(3-1)} band. A linear least square fit gives us this relation with
$\rm T_{eff} =4974.85 - 56.53 \times CO(3-1)$  with an r.m.s of 120\,K which is slightly higher as for $^{12}$CO(2-0) .  However, both CO band heads are clearly excellent temperature indicators. In addition, we tried to measure the $^{12}$CO(4-2) band  centered at $\rm \lambda_{c} = 2.3535\,\mu m$.  Although we see clearly, as for the other bands, an increase of the band strength with decreasing temperature, the intrinsic scatter is much higher ($\sim 200\,K$). For that reason we do not discuss further this relation.

How does the spectroscopic temperatures compare with those derived from photometric measurements?

\begin{figure}[!htbp]
\includegraphics[width=0.4\textwidth]{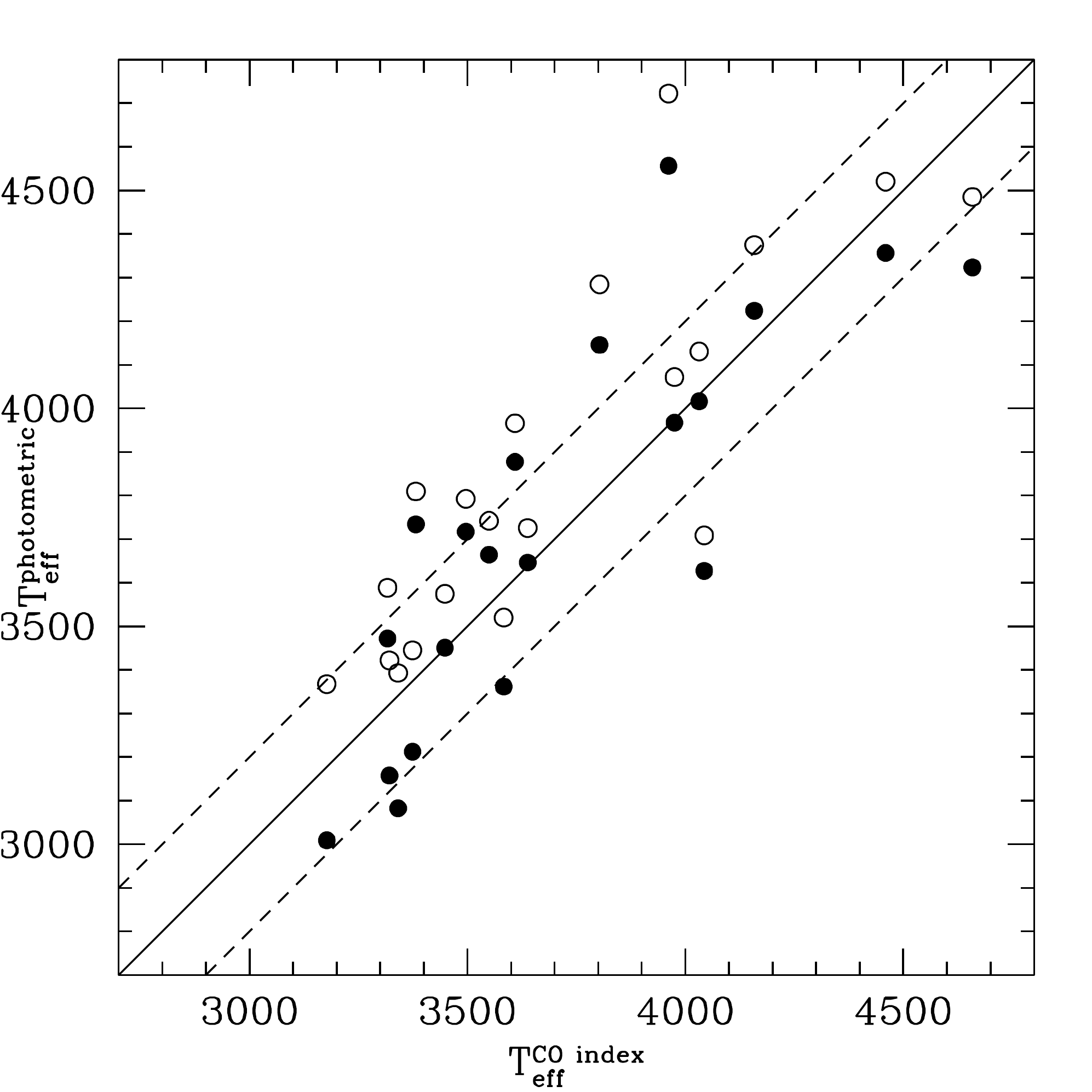}
\caption{Photometric temperatures vs. spectroscopically derived temperatures. Open circles use the relation by Houdashelt et al. (2000), black circles those from Montegriffo et al. (1998).}
\label{TeffspvsTeffphot}
\end{figure}

Figure~\ref{TeffspvsTeffphot} shows the comparison between temperatures derived from photometric colours and from spectroscopy for our sample (see Table.~1) In order to calculate the photometric temperatures, we used here the temperature scale of M giants from \citet{montegriffo1998} which is derived from black-body fits of population II stars  and that of \citet{houdashelt2000} using synthetic colors of M giants from MARCS models.  We used the dereddened $\rm (J-K)_{0}$  colour of our stars to obtain the photometric temperatures. The mean difference between \citet{montegriffo1998} and the  temperatures from
the $^{12}$CO(2-0) index  is $\rm 10\,K \pm 200$ while for the the photometric temperatures of  \citet{houdashelt2000} they are about 90\,K systematically  higher  with a  higher r.m.s dispersion $\rm \pm 250\,K$.

 %If one would add  the additional 28 stars  from \citet{ryde2015}, the dispersion increases slightly which is expected since the photometric temperatures are more uncertain for the inner fields of the bulge.  Note that for the galactic Center itself,  no photometric temperatures are available as they do not have  J counterparts.

%The majority of the stars show differences less than 200\,K (indictated by the dashed lines in Fig.~\ref{TeffspvsTeffphot} between spectroscopic and photometric temperatures.
%However,  the photometric temperatures of  \citet{houdashelt2000} are systematically 100\,K   higher compared to the spectroscopic ones.

%Bessell et al. (1998) provided a relation for (V − K) vs. BCK derived from model spectra down to Teff = 3600 K. In the same year, Montegriffo et al. (1998) published a set of bolometric corrections derived from black-body fits to population II stars. 

%\subsection{Surface gravity}

%\citet{ramirez2000} pointed out that the combination of the NaI, CaI lines and the CO band can be used as luminosity
%discriminator to separate e.g. dwarfs from giants or supergiants.  This results has been confirmed by  \citet{ivanov2004}.
%Figure~\ref{Ramirezlogg} shows the surface gravity as a function of
%the combined lines of NaI, CaI and the CO band. Unfortunately, only for 8 stars  of our sample  do have known surface gravities. The rest of the stars
%have only approximate values set all to log\,g=0.5 (see Table.~1).  
%We see a small trend in the sense that the
%gravity index  increases with decreasing $\rm log{g}$ but clearly more stars are needed to confirm this tendancy  if one can use this spectral index
 %to derive surface gravities
%\begin{figure}[!htbp]
%\includegraphics[width=0.5\textwidth]{Teff_loggRamirez.pdf}
%\caption{Logg vs. surface  gravity index of Ramirez et al. (2000)}
%\label{Ramirezlogg}
%\end{figure}

\subsection{Metallicity}

With the upcoming low-resolution multi-fiber near-IR spectrographs (MOONS,MOS, etc..),  the main question arises how accurate one
can get metallicities with low-resolution infrared  spectra compared to high-resolution IR spectra.  During the last years several studies came out
to trace metallicity distribution functions as well as metallicity gradients. \cite{gonzalez2011} derived photometric metallicities based on 
VVV data  along the bulge minor axis, where  accurate high-resolution spectroscopic metallicities are available. They found a remarkable agreement between
both methodes. \citet{gonzalez2013} obtained  a full low-resolution metallicity map over the VVV Bulge footprint where they revealed a clear metallicity gradient of 0.28\,dex/kpc. Due to the high extinction, this map is restricted to $\rm |b| > 3^{o}$.  
\citet{ramirez1997} have established an [Fe/H] scale for Galactic globular clusters based on medium-resolution (1500–3000) infrared K-band spectra. The technique uses the same absorption features we use here: NaI, Ca, and CO and was calibrated on globular cluster stars. The technique was calibrated and tested for globular cluster giants with $\rm -1.8 < [Fe/H] < -0.1$ and $\rm -7 < M_{K} < -4$ and has a typical uncertainty of $\rm \pm 0.1 dex.$  \citet{ramirez2000} calculated  metallicities for 110 M giants in the Inner Bulge  using these spectral indices of  NaI, Ca and CO.  \citet{schultheis2003} applied this technique to get the metallicity
distribution of ISOGAL sources in the inner degree of our Galaxy.  \citet{do2015} observed low-resolution K-band spectra ($\rm R \sim 5000$) of late-type giants within the central 1\,pc and obtained metallicities by fitting very sparsely sampled synthetic spectra, intended for calculating MARCS model atmospheres (which should actually not be used as synthetic stellar spectra; see the MARCS homepage; www.marcs.astro.uu.se). Their estimated uncertainties are consequently large, larger than 0.3 dex, with some very uncertain systematic effects leading to extremely metal-rich stars.

\begin{figure}[!htbp]
\includegraphics[width=0.4\textwidth]{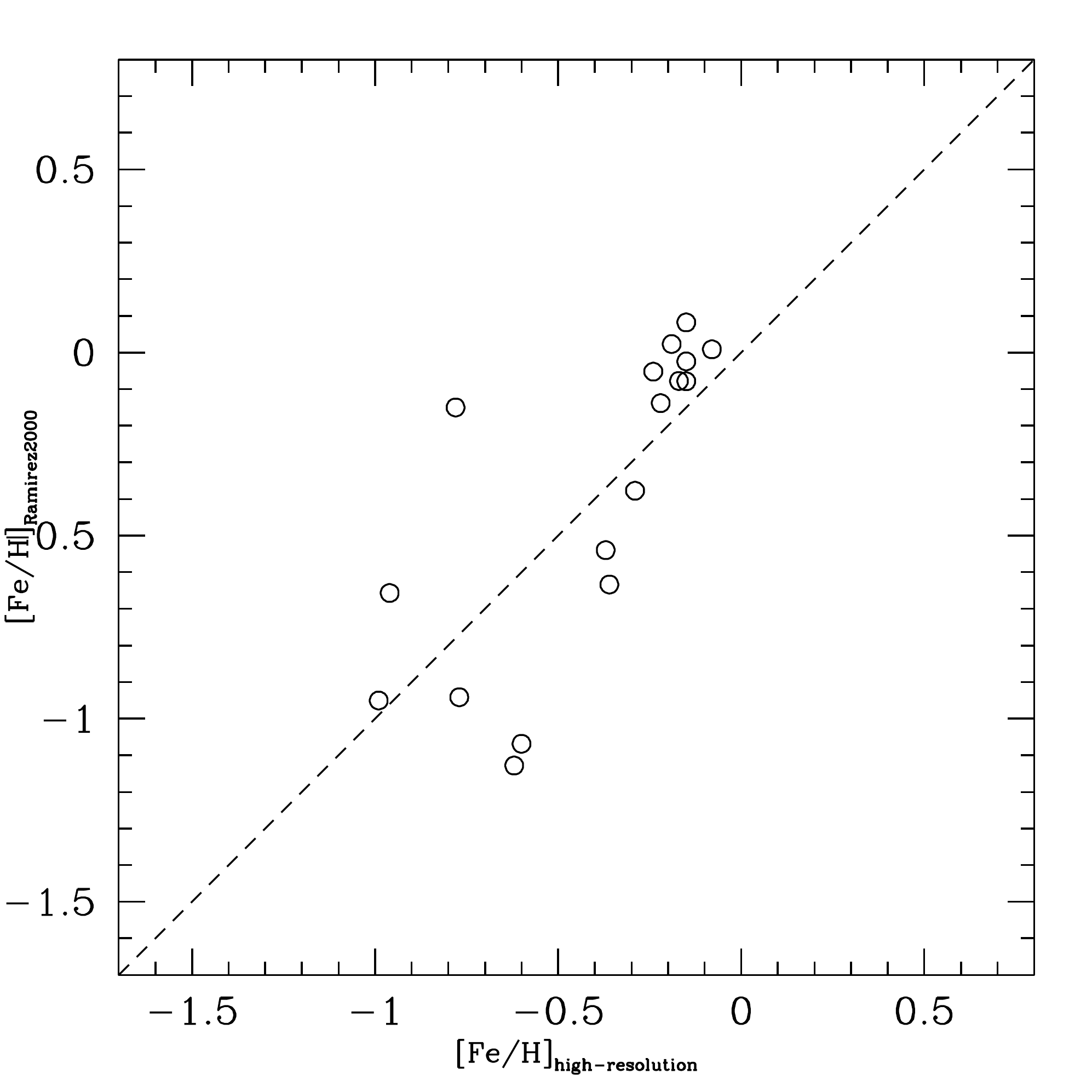}
\caption{[Fe/H] comparison between low-resolution spectra (Ramirez et al. 2000) and high resolution spectra from our calibration sample (Table 1).}
\label{metallicities}
\end{figure}

We used solution 1 from \citet{ramirez2000}  which does not take any photometric quantities into account with the  following relation:
%\begin{equation}

$\rm  [Fe/H] = -1.782 + 0.352 \times EW(Na) - 0.0231 \times EW(Na)^{2}- 0.0267 \times EW(Ca) + 0.0129 \times EW(Ca)^{2} +0.0472 \times EW(CO) - 0.00109 \times EW(CO)^{2} $

%\end{equation} 

%\citet{ryde2015} and \citet{ryde2015a}.

where EW(CO), EW(Na) and EW(Ca) are the equilavent widths of CO, Na and Ca, respectively.  Figure~\ref{metallicities}  shows the  comparison of
metallicities derived as in \citet{ramirez2000}  from low-resolution spectra and those  from high-resolution infrared spectra coming from our calibration sample.
One can notice  that there is a general correlation between the metallicities
derived from high-resolution IR spectra and those from low-resolution spectra. The  dispersion is in the order of $\rm \sim 0.2\,dex$.
 %(ii) We see an increased spread or  even a  plateau at about $\rm [Fe/H]=0.0$ for
%the low-resolution part where the metal-rich stars do not follow this relation anymore. This indicates that the Na and the Ca lines starts to saturate for metal-rich stars. Indeed, by tracing
%the equilavent width of NaI and CaI alone, one see this plateau at around solar metallicity. This means
% that the proposed relation of \citet{ramirez2000} is only valid for M giants with sub-solar metallicities. Therefore this metallicity index is not suited for galactic Bulge studies or at least for the metal-rich component of the metallicity distribution of bulge stars but can be used in
%low-metallicity systems such as globular clusters or dwarf spheroidals.

\section{Application of the method to  M giants in the galactic Bulge}

\subsection{Temperature}

\begin{figure}[!htbp]
\includegraphics[width=0.4\textwidth]{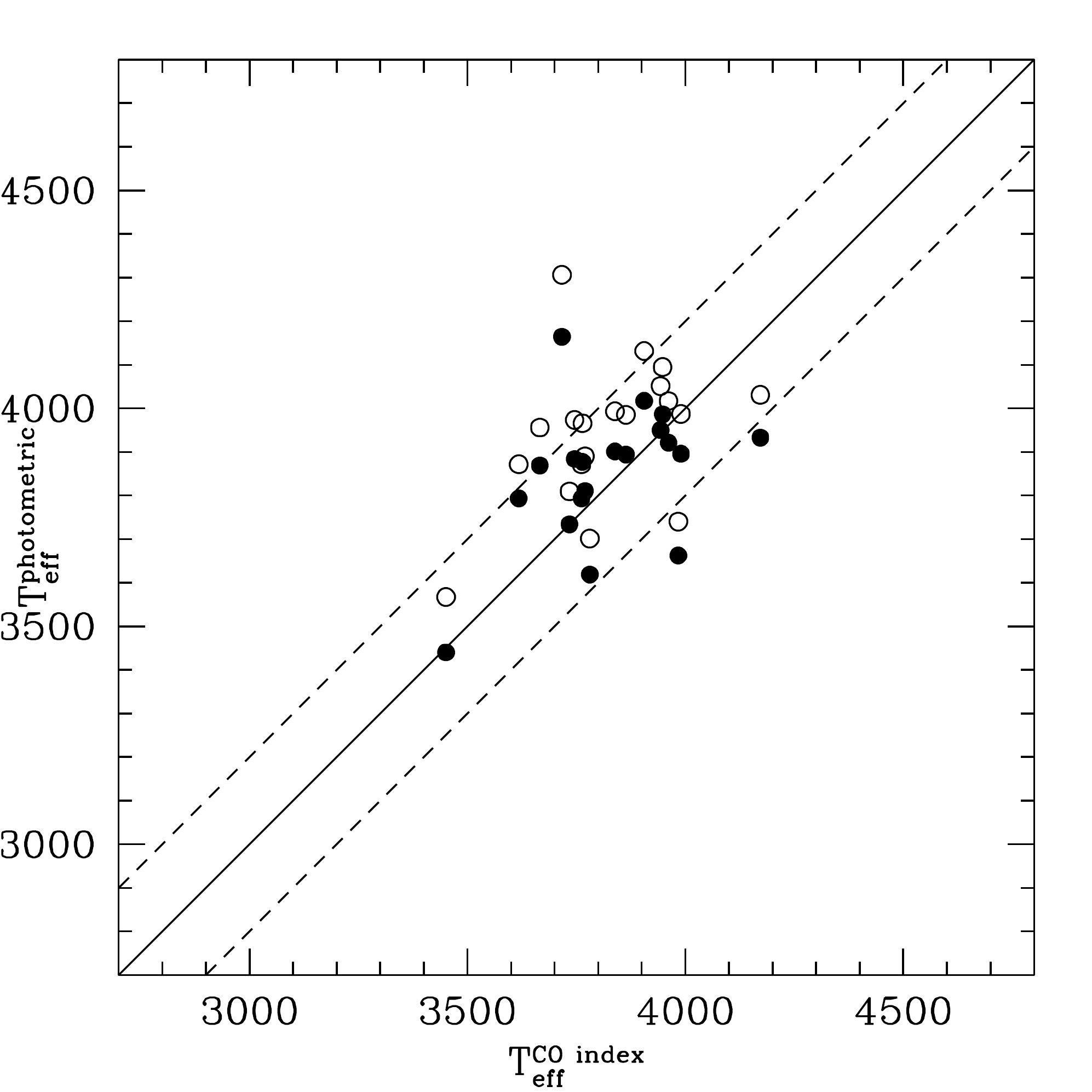}
\caption{Photometric temperatures vs. spectroscopically derived temperatures for the stars of \citet{ryde2015a}. Open circles use the relation by Houdashelt et al. (2000), black circles those from Montegriffo et al. (1998). Due to the missing J counterpart, the stars in te galactic Center (see Table~2) were omitted.}
\label{TeffspvsTeffphot1}
\end{figure}

We applied this method to the M giants sample of \citet{ryde2015a} located in the inner 300\,pc
of the Milky Way. Figure~\ref{TeffspvsTeffphot1}  shows the comparison between our method and photometrically derived temperatures.  We see  here a more scattered diagram
 which is due to variable, patchy extinction making the photometric temperatures less reliable. Due to the
extreme high extinction in the galactic Center region, the M giants of \citet{ryde2015} do not have J counterparts and are not included in Fig.~\ref{TeffspvsTeffphot1}

\subsection{Metallicity}
\begin{figure}[!htbp]
\includegraphics[width=0.4\textwidth]{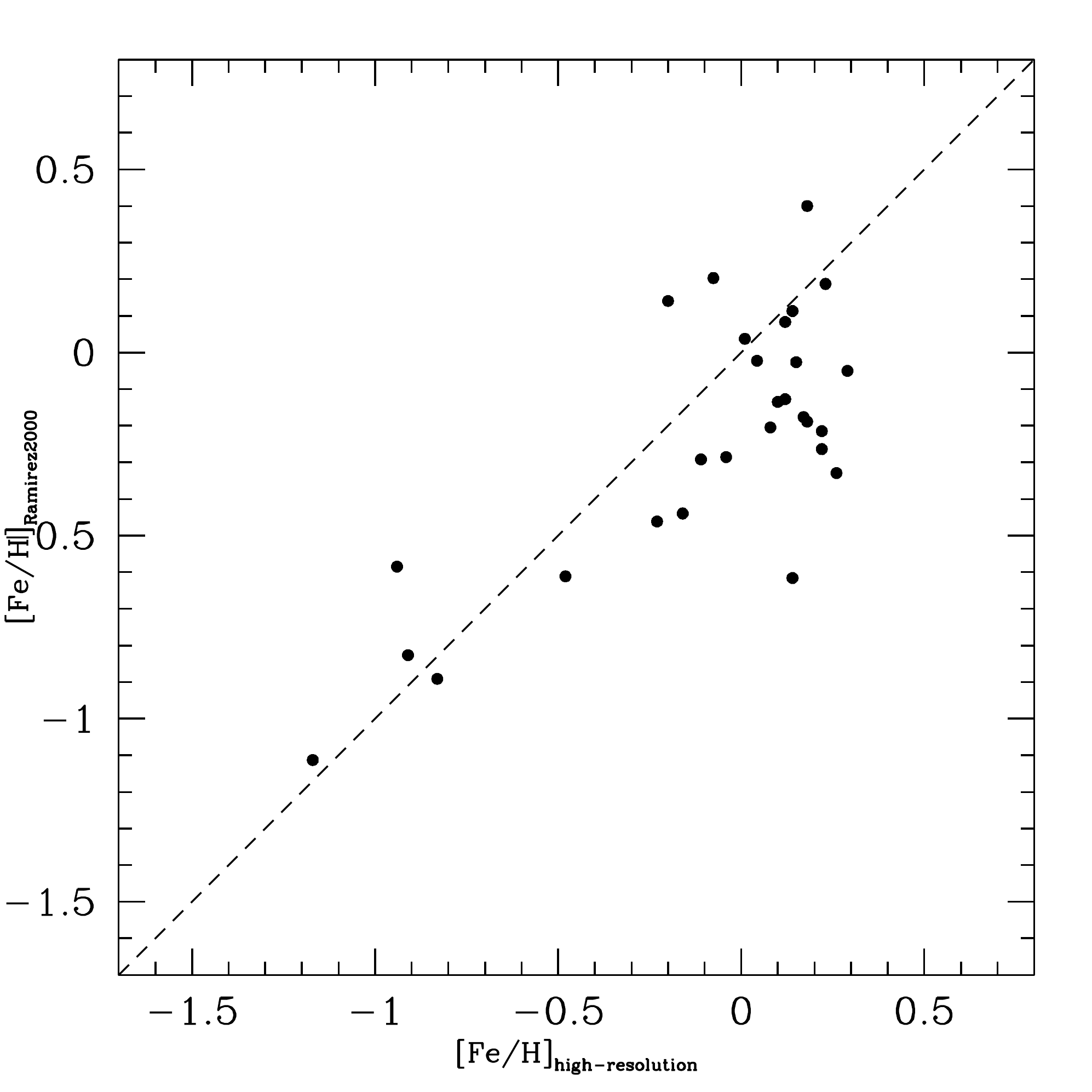}
\caption{[Fe/H] comparison between low-resolution spectra and high resolution spectra for the stars from \citet{ryde2015} and \citet{ryde2015a}. }
\label{metallicities1}
\end{figure}

Figure~\ref{metallicities1}  shows the application of this method to the M giant Bulge sample of \citet{ryde2015} and \citet{ryde2015a}. We see an increased spread or  even a  plateau at about $\rm [Fe/H]=0.0$ for the low-resolution part where the metal-rich stars do not follow this relation anymore. This indicates that the Na and the Ca lines saturate in the most metal-rich stars. Indeed, by tracing
the equilavent width of NaI and CaI alone, one see this plateau at around solar metallicity. This means  that the proposed relation of \citet{ramirez2000} is only valid for M giants with sub-solar metallicities. Therefore this metallicity index is not suited for galactic Bulge studies or at least for the metal-rich component of the metallicity distribution of bulge stars but can be used in
low-metallicity systems such as globular clusters or dwarf spheroidals. 

%\section{Discussion} \label{Discussion}

\section{Conclusions} \label{Conclusions}

In high extincted regions such as the galactic Bulge and especially the galactic Center, the precision in the  photometric temperatures and gravities are hampered
 by  the poor knowledge of interstellar extinction.
We have studied  20 well-known M giants with well-known stellar parameters and covering the temperature range $\rm 3200 < T_{eff} < 4500\,K$ and with a metallicity
range $\rm -1.2 < [Fe/H] < 0.5$.  We confirm the straight relation found by \citet{blum2003} between $\rm T_{eff}$ and the $^{12}$CO(2-0) index with a dispersion of 95\,K. We also find a relation between  $^{12}$CO(3-1) and $\rm T_{eff}$  with a dispersion of 120\,K.
We do not find  any critical dependence of these relations on  metallicity or on the adopted spectral resolution, which makes them a very powerful tool to obtain accurate temperatures in the Inner galactic Bulge and the
 galactic Center which has been applied by \cite{ryde2015} and \citet{ryde2015a}. 
Finally, we confirm that the combination of NaI, CaI and CO band is an  excellent metallicity index as pointed out by \citet{ramirez2000}. However, this relation is only valid
for M giants with sub-solar metallicities but  can not be applied for metal-rich stars and thus for the metal-rich part of the  galactic Bulge.

\begin{acknowledgements}
We want to thank the  referee L. Origlia for her extremly useful comments and suggestions.
N.R. acknowledges support from the Swedish Research Council, VR (project number 621-2014-5640), and Funds from Kungl. Fysiografiska S\"allskapet i Lund.
(Stiftelsen Walter Gyllenbergs fond and M\"arta och Erik Holmbergs donation).

\end{acknowledgements}

\bibliographystyle{aa}
\bibliography{lowresolution_accepted}

\end{document}